\documentclass{mem}
\usepackage{natbib}\usepackage{txfonts}\usepackage{balance}
\usepackage{graphicx}
\usepackage{wrapfig}
\usepackage{multicol}
\usepackage{lipsum}
\usepackage{tikz}
\usepackage{float}
\usepackage[section]{placeins}
\usepackage{stfloats}
\usepackage[unskipbreak]{cuted}
\usepackage[font=small,labelfont=bf]{caption}
\usepackage{txfonts}
\usepackage{indentfirst}

\input{insbox}

\idline{75}{282}
\begin{document}
\def\teff{$T\rm_{eff }$}
\def\kms{$\mathrm {km s}^{-1}$}

\title{
The Transient High-Energy Sky and Early Universe Surveyor (THESEUS)
}

   \subtitle{}

\author{
L. \,Amati\inst{1} 
\and E. \,Bozzo\inst{2}
\and P. \,O'Brien\inst{3} 
\and D. \,G\"otz\inst{4} \\
(on behalf of the THESEUS consortium)
          }

\institute{
INAF-OAS Bologna, via P. Gobetti 101, I-40129 Bologna, Italy \\
\email{lorenzo.amati@inaf.it}
\and
Department of Astronomy, University of Geneva, chemin d'Ecogia 16, 1290 Versoix, Switzerland
\and
Department of Physics and Astronomy, University of Leicester, Leicester LE1 7RH, United Kingdom
\and
CEA, Université Paris-Saclay, F-91191 Gif-sur-Yvette, France 
}

\authorrunning{L. Amati}

\titlerunning{THESEUS}

\abstract{The Transient High-Energy Sky and Early Universe Surveyor (THESEUS) is a mission concept developed in the last years by a large European consortium and currently under study by the European Space Agency (ESA) as one of the three candidates for next M5 mission (launch in 2032). THESEUS aims at exploiting high-redshift GRBs for getting unique clues to the early Universe and, being an unprecedentedly powerful machine for the detection, accurate location (down to $\sim$arcsec) and redshift determination of all types of GRBs (long, short, high-z, under-luminous, ultra-long) and many other classes of transient sources and phenomena, at providing a substantial contribution to multi-messenger time-domain astrophysics. Under these respects, THESEUS will show a strong synergy with the large observing facilities of the future, like E-ELT, TMT, SKA, CTA, ATHENA, in the electromagnetic domain, as well as with next-generation gravitational-waves and neutrino detectors, thus greatly enhancing their scientific return. 
}
\maketitle{}

\section{Introduction}

The main feature of the modern Astrophysics is the rapid development of multi-messenger astronomy. At the same time, relevant open issues still affect our understanding of the cosmological epoch (a few millions years after the ``big-bang'') at which first stars and galaxies start illuminating the Universe and re-ionizing the inter-galactic medium. 

In this context, a substantial contribution is expected from the Transient High Energy Sky and Early Universe Surveyor (THESEUS\footnote{https://www.isdc.unige.ch/theseus}), a space mission concept developed by a large European consortium including Italy, UK, France, Germany, Switzerland, Spain, Poland, Denmark, Czech Republic, Ireland, Hungary, Slovenia, ESA, with Lorenzo Amati (INAF-OAS Bologna, Italy) as a lead proposer. In May 2018, THESEUS was selected by ESA for a Phase 0/A study as one of the three candidates for the M5 mission within the Cosmic Vision programme. The end of Phase A and a down-selection to one mission to be implemented is expected for mid-2021. The launch of the selected M5 mission is planned for 2032. Details on the THESEUS science objectives, mission concept, and expected performances are reported in \citet{amati18} and \citet{stratta18}. 
\begin{figure}[h]
\begin{center}
\includegraphics[width=6.5cm]{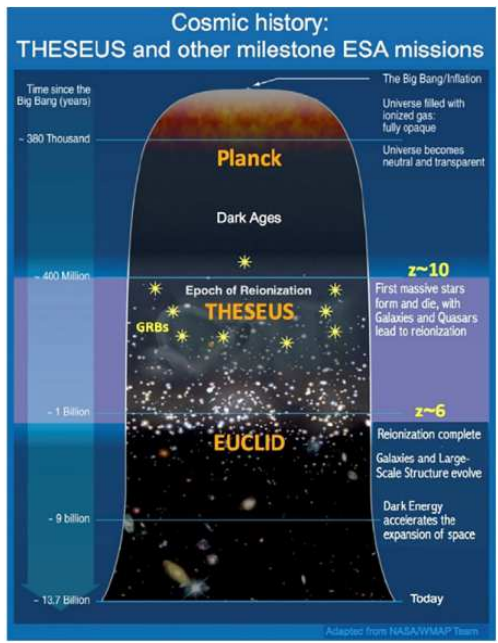}
\end{center}
\caption{Gamma-Ray Bursts in the cosmological context and the role of THESEUS (adapted from a picture by the NASA/WMAP Science team).}
\label{fig:theseus}
\end{figure}

\section{Scientific objectives}

THESEUS is designed to vastly increase the discovery space of high energy transient phenomena over the entirety of cosmic history (see Fig.~\ref{fig:theseus}).
\begin{figure*}[t!]
\begin{center}
\includegraphics[width=6.0cm]{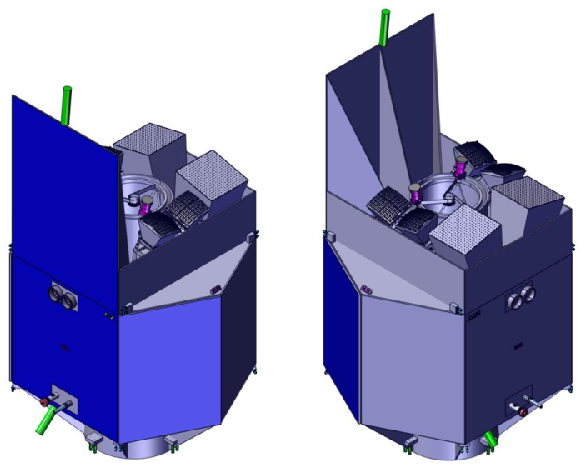}
\includegraphics[width=7.0cm]{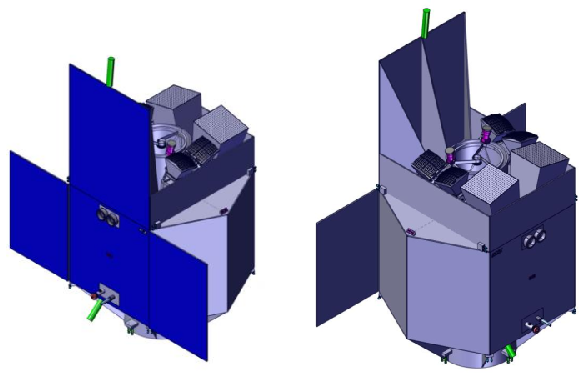}
\end{center}
\caption{Sketch of the THESEUS spacecraft and payload accommodation (left stowed configuration, right deployed configuration. The IRT is placed in the middle of the optical bench, and are clearly visible the 4 SXI squared cameras, as well as the two XGIS rectangular cameras (credits: ESA).}
\label{fig:sc}
\end{figure*}
\begin{figure}[t!]
\begin{center}
\includegraphics[width=3.5cm]{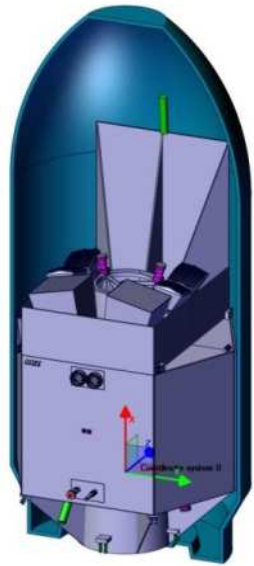}
\end{center}
\caption{Sketch of THESEUS inside the VEGA fairing (credits: ESA).}
\label{fig:vega}
\end{figure}

Because of their huge luminosities, mostly emitted in the X and gamma-rays, their redshift distribution extending at least to z$\sim$9 and their association with explosive death of massive stars and star forming regions, GRBs are unique and powerful tools for investigating the early Universe: SFR evolution, physics of re-ionization, galaxies metallicity evolution and luminosity function, first generation (pop III) stars. THESEUS will obtain a statistical sample of high–z GRBs, which in turns allow us to \citep{amati18}:
\begin{itemize}
\item measure independently the cosmic star–formation rate, even beyond the limits of current and future galaxy surveys;
\item directly (or indirectly) detect the first population of stars (pop III);
\item obtain the number density and properties of low-mass galaxies (even JWST and ELTs surveys will be not able to probe the faint end of the galaxy Luminosity Function at z$>$8-10);
\item evaluate the neutral hydrogen fraction;
\item measure the escape fraction of UV photons from high-z galaxies;
\item study the early metallicity of the ISM and IGM and its evolution. 
 \end{itemize}
 
Through a carefull design optimization, a mission capable of substantially increase the rate of identification and characterization of high-z GRBs can also provide a survey of the high-energy sky from soft X-rays to gamma-rays with an unprecedented combination of wide Field Of View (FoV), source location accuracy and sensitivity below 10~keV. For this reason, THESEUS will also provide a substantial contribution also to time-domain Astrophysics, in general, and in particular to the newly born and fastly growing field of multi-messenger Astrophysics. For instance, THESEUS will be able to provide detection, accurate location, characterization and possibly redshift measurement of electromagnetic emission (short GRBs, possible soft X-ray transient emission, kilonova emission in the near-infrared) from gravitational-wave sources like NS-NS or NS-blackhole (BH) mergers \citep{stratta18}.  

THESEUS will be an unprecedentedly powerful machine for the detection, accurate localization (down to $\sim$arcsec) and redshift determination of all types of GRBs (long, short, high-z, under-luminous, ultra-long) and many other classes of transient sources and phenomena. THESEUS will also provide a substantial contribution to multi-messenger time-domain astrophysics. The mission capabilities in exploring the multi-messenger transient sky can be summarized as follow: 
\begin{itemize}
\item Localize and identify the electromagnetic counterparts to sources of gravitational radiation and neutrinos, which may be routinely detected in the late '20s / early '30s by next generation facilities like aLIGO/aVirgo, eLISA, ET, or Km3NET;
\item Provide real-time triggers and accurate ($\sim$1 arcmin within a few seconds, $\sim$1~arcsec within a few minutes) localizations of both high-energy transients for follow-up with next-generation optical-NIR (E-ELT, JWST if still operating), radio (SKA), X-rays (ATHENA), TeV (CTA) telescopes, as well as to LSST sources; 
\item Provide a fundamental step forward in the comprehension of the physics of various classes of transients and fill the present gap in the discovery space of new classes of transient events.
\end{itemize}
\begin{figure*}[t!]
\begin{center}
\includegraphics[width=4.5cm]{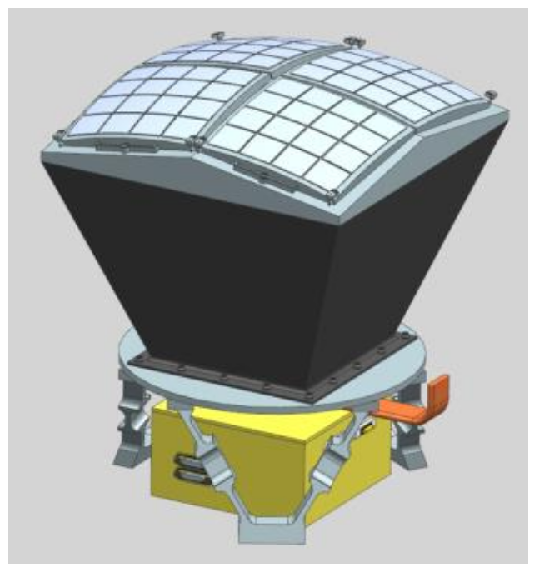}
\includegraphics[width=8.0cm]{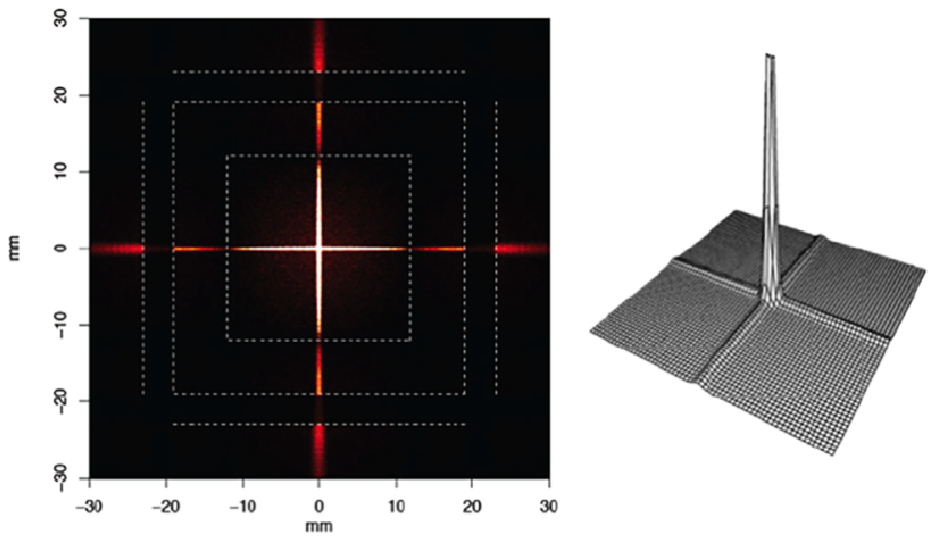}
\end{center}
\caption{Left: sketch of one SXI camera. Right: the SXI point spread function.}
\label{fig:sxi}
\end{figure*}

In the field of gravitational wave source, THESEUS capabilities will permit us to: 
\begin{itemize}
\item detect short GRBs over a large FoV with arcmin localization; 
\item detect the Kilonovae and provide their arcsec localization and characterization; 
\item possibly detect weaker isotropic X-ray emissions. 
 \end{itemize}

\section{Mission concept and payload}

THESEUS will be capable to achieve the exceptional scientific objectives summarized above thanks to a smart combination of instrumentation and mission profile. The mission will carry on-board two large FoV monitors covering simultaneously a 1~sr FoV in the soft X-rays (0.3-5~keV) with unprecedented sensitivity and arcmin location accuracy  and several sr FoV from 2~keV up to 20~MeV, with additional source location capabilities of a few arcmin from 2 to 30 keV. Once a GRB or a transient of interest is detected by one or both the monitors, the THESEUS spacecraft will autonomously slew quickly to point, within a few minutes, an on-board near infra-red telescope (70~cm class operating from 0.7 to 1.8~$\mu$m) toward the direction of the transient, so to catch the fading NIR afterglow or, e.g., the kilonova emission, localizing it at a  1~arcsec accuracy and measuring its redshift through photometry and moderate resoluton spectroscopy. 

The detailed description of THESEUS can be found in \citet{amati18}. A scketch of the THESEUS spacecraft, showing also the accommodation of all payload elements, is presented in Fig.~\ref{fig:sc}. This sketch is the result of the Concurrent Design Facility (CDF) study performed by ESA at the end of the phase 0 before the kick-off of phase A (end of 2018). It is expected that THESEUS will be injected into a low equatorial orbit ($<$6~deg. inclination, $\sim$600~km altitude) with a VEGA-C launcher (see Fig.~\ref{fig:vega}). We provide a short description of the THESEUS payload, comprising the SXI, XGIS, and the IRT instruments, in the following sub-sections.

\subsection{The Soft X-ray Imager}

The soft X-ray Imager (SXI) comprises a set of four sensitive lobster-eye telescopes observing in the 0.3-5~keV energy band and providing a FoV of $\sim$1sr. The expected source location accuracy is 0.5-1~arcmin. A few details of the instrument are provided in Fig.~\ref{fig:sxi}. The SXI is being developed by a UK-led consortium.

\subsection{The X- and Gamma-ray imaging spectrometer}

\begin{figure*}[t!]
\begin{center}
\includegraphics[width=9.0cm]{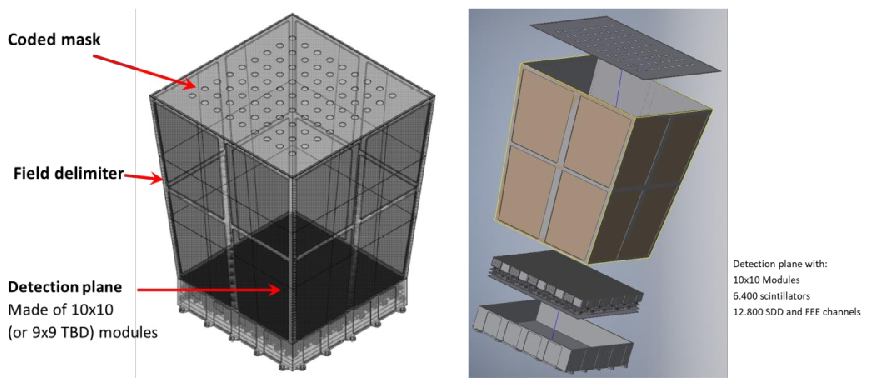}
\includegraphics[width=9.0cm]{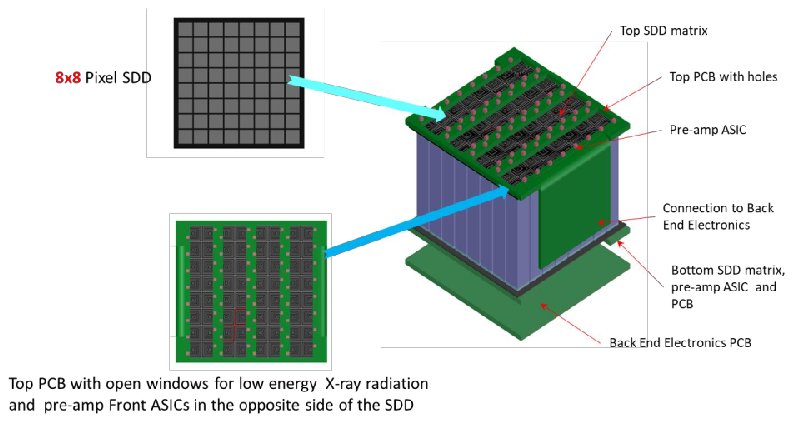}
\end{center}
\caption{Top: sketch of one XGIS camera. Bottom: details of one of the 100 modules comprised within the focal plane of each XGIS camera.}
\label{fig:xgis}
\end{figure*}

\begin{figure}[t!]
\begin{center}
\includegraphics[width=6.0cm]{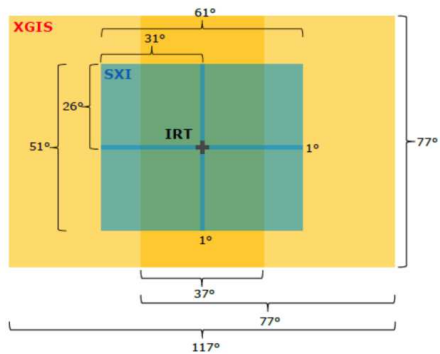}
\end{center}
\caption{Combined FoV of all THESEUS instruments (credits: ESA).}
\label{fig:fov}
\end{figure}

The X- and Gamma-rays Imaging Spectrometer (XGIS) comprises 2 coded-mask cameras using bars of Silicon diodes coupled with CsI crystal scintillators (see Fig.~\ref{fig:xgis}). The instrument is operating in the 2~keV-10~MeV energy band, providing imaging capabilities only in the 2-30~keV energy range. It operates as a collimated instrument between 30-150~keV and as an all-sky monitor at higher energies. Depending on the operating mode, the XGIS can achieve a FoV as large as $\sim$2-4~sr and provides a source location accouracy of about 5~arcmin. In order to optimize the detection of high energy transients and GRBs in particular, the FoV of the XGIS partly overlap with that of the SXI (see Fig.~\ref{fig:fov}). The XGIS is being developed by an Italian-led consortium. 

\subsection{The InfraRed Telescope}

\begin{figure}[t!]
\begin{center}
\includegraphics[width=6.5cm]{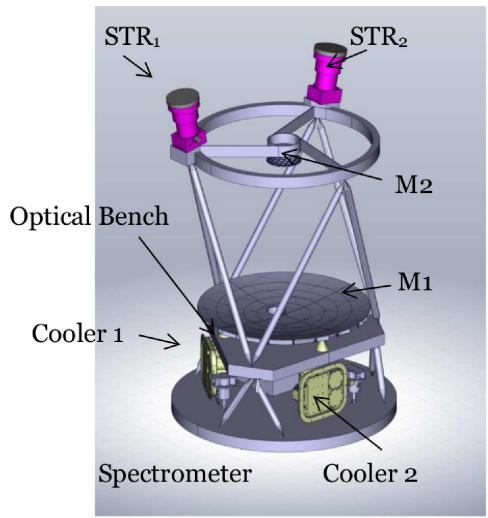}
\end{center}
\caption{The IRT assembly (credits: ESA).}
\label{fig:irt}
\end{figure}

The InfraRed Telescope (IRT) is a 0.7m class IR telescope operating between 0.7-1.8~$\mu$m. A design based on a off-axis Korsch model is presently (see Fig.~\ref{fig:irt}) assumed, resulting in a FoV of 15$\times$15~arcmin and providing both imaging and moderate resolution spectroscopy capabilities (up to R=500). The IRT is being developed by a French-led consortium.

\section{THESEUS performances}

\subsection{The early Universe with GRBs}

\begin{figure}[t!]
\begin{center}
\includegraphics[width=6.5cm]{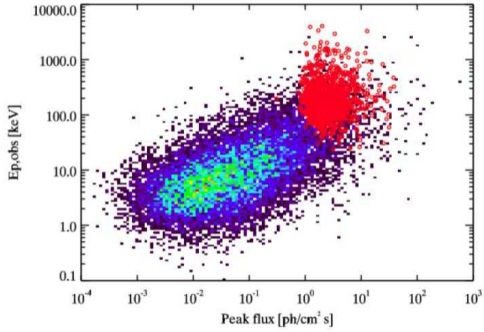}
\end{center}
\caption{GRB distribution in the peak flux - spectral peak energy (E$_{\rm p}$) plane according to most recent population synthesis models and measurements \citep{amati18}. For all shown GRBs, THESEUS will be able to provide detection, accurate location, characterization and a measurement of the redshift. The low-E$_{\rm p}$ - low peak flux region is populated by high-redhsift GRBs (shown in dark blue, blue, ligt blue, green, yellow), a population unaccessible by current facilities, while the high E$_{\rm p}$ region highlighted with red points shows the region where most short GRBs will lay. }
\label{fig:grb}
\end{figure}

THESEUS will have the ideal combination of instrumentation and mission profile for detecting all types of GRBs (long, short/hard, weak/soft, high-redshift), localizing them from a few arcmin down to arsec and measure the redshift for a large fraction of them (see Fig.~\ref{fig:grb}).

In addition to the GRB prompt emission, THESEUS will also detect and localize down to 0.5-1~arcmin the soft X-ray short/long GRB afterglows, of NS-NS (BH) mergers and of  many classes of galactic and extra-galactic transients. For several of these sources, the IRT will provide a characterization of the associated IR counterpart, including a location within 1~arcsec and, possibly, the redshift.

\begin{figure}[t!]
\begin{center}
\includegraphics[width=6.5cm]{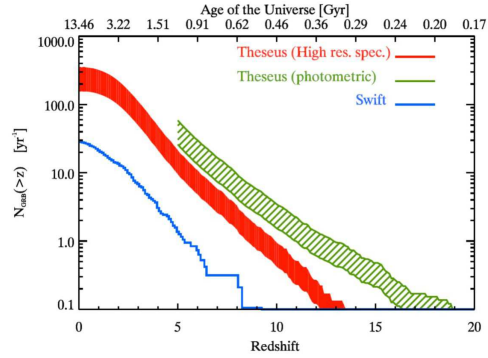}
\end{center}
\caption{The yearly cumulative distribution of GRBs with redshift determination as a function of redshift for Swift and THESEUS. We note that these predictions are conservative in so far, as they reproduce the current GRB rate as a function of redshift. However, thanks to its improved sensitivity, THESEUS can detect a GRB of $E_{\rm iso}$$=$10$^{53}$~erg (corresponding to the median of the GRB radiated energy distribution) up to $z=12$. Indeed, our currently poor knowledge of the GRB rate - star formation rate connection does not preclude the existence of a sizable number of GRBs at such high redshifts, in agreement with recent expectations on Pop III stars.}
\label{fig:grb2}
\end{figure}

The impact of the THESEUS measurements for shedding light on the study of the early Universe exploiting GRBs is represented in Fig.~\ref{fig:grb2}, where we show the expected number per year of GRBs detected, localized, and for which a redshift measurement is achieved. The THESEUS expected performance is compared to the present situation, which is the result of a large coordinated effort between Swift, Konus-WIND, Fermi/GBM, and several on-ground robotic/large telescopes. 

\subsection{Multi-messenger and time-domain astrophysics}

As anticipated in few of the previous sections, THESEUS will be capable of monitoring the electro-magnetic (EM) domain a number of different expected gravitational wave source counterparts, including: 
\begin{itemize}
\item NS-NS / NS-BH mergers: for these events, THESEUS is expected to detect the collimated EM emission from short GRBs, as well as their afterglows (the currently estimated event rate is of $\lesssim$1~yr$^{-1}$ for the GW detectors of the second generation but up to $\sim$20~yr$^{-1}$ for the third generation detectors as the Einstein Telescope). THESEUS is also expected to detect the NIR and soft X-ray isotropic emissions from macronovae, as well as from off-axis afterglows and, for NS-NS, identify newly born magnetar spinning down in the millisecond domain (the rate of GW detectable NS-NS or NS-BH systems is estimated at dozens-hundreds~yr$^{-1}$). 
\item Core collapse of massive stars: for these events, THESEUS is expected to detect the emission from long GRBs, LLGRBs, as well as ccSNe (in these cases the predictions on the energy released in GWs is much more uncertain and the estimated rate of events is of $\sim$1~yr$^{-1}$). 
\item Flares from isolated NSs: for these events, THESEUS is expected to be able to detect the typical emission from, e.g, the Soft Gamma Repeaters (although the associated GW energy content is estimated to be only $\sim$0.01\%-1\% of the EM emission). 
\end{itemize}

THESEUS will be able to detect, localize, characterize and measure the redshift for NS-NS / NS-BH mergers thorugh the following channels:
\begin{itemize}
\item collimated on-axis and off-axis prompt gamma-ray emission from short GRBs; 
\item NIR and soft X-ray isotropic emissions from kilonovae, off-axis afterglows and, for NS-NS, from newly born ms magnetar spindown.
\end{itemize}

THESEUS will thus beautifully complement the capabilities of next generation GW detectors (e.g., Einstein Telescope, Cosmic Explorer, further advanced LIGO and Virgo, KAGRA, etc.) by promptly and accurately localizing e.m. counterparts to GW signals form NS-NS and NS-BH mergers and measuring their redshift. These combined measurements will provide unique clues on the nature of the progenitors, on the extreme physics of the emission and, by exploiting simultaneous redshift (from e.m. counterpart) and luminosity distance (form GW signal modeling) of tens of sources, fully exploit the potentialities of multi-messenger Astrophysics for cosmology.

\subsection{Time domain Astronomy and GRB physics}

The unique capabilities of THESEUS, will also allow us to provide relevant contributions to the more general field of time-domain Astronomy and, of course, to GRB science. As a few exaomples, THESEUS will provide the astrophysical community with:
\begin{itemize}
\item survey capabilities of transient  phenomena similar to the Large Synoptic Survey Telescope (LSST) in the optical: a remarkable scientific sinergy can be anticipated; 
\item substantially increased detection rate and characterization of sub-energetic GRBs and X-Ray Flashes; 
\item unprecedented insights in the physics and progenitors of GRBs and their connection with peculiar core-collapse SNe.
\end{itemize}
    
\section{Conclusions}

THESEUS, under study by ESA and a large European collaboration with strong interest by international partners (e.g., US) will fully exploit GRBs as powerful and unique tools to investigate the early Universe and will provide us with unprecedented clues to GRB physics and sub-classes. This mission will also play a fundamental role for GW/multi-messenger and time domain astrophysics at the end of next decade, also by providing a flexible follow-up observatory for fast transient events with multi-wavelength ToO capabilities. THESEUS observations will thus impact on several fields of Astrophysics, Cosmology and even fundamental Physics and will enhance importantly the scientific return of next generation multi messenger (aLIGO/aVirgo, LISA, ET, or Km3NET;) and e.m. facilities (e.g., LSST, E-ELT, SKA, CTA, ATHENA)

In addition, THESEUS scientific return will include significant Observatory Science, e.g. studying thousands of faint to bright X-ray sources through the unique simultaneous availability of broad band X-ray and NIR observations. 

THESEUS will be a really unique and superbly capable facility, one that will do amazing science on its own, but also will add huge value to the currently planned new photon and multi-messenger Astrophysics infrastructures in the 2020s to $>$2030s.

\bibliographystyle{aa}

\end{document}